\documentclass[twocolumn, showpacs, preprintnumbers, amsmath, amssymb]{revtex4}

\usepackage{graphicx}
\usepackage{dcolumn}
\usepackage{bm}
\usepackage{mathbbol}
\usepackage{epsf}
\usepackage{mathrsfs}
\usepackage{slashed}
\usepackage{stmaryrd}

\newcommand{\be}{\begin{equation}}
\newcommand{\ee}{\end{equation}}
\newcommand{\bea}{\begin{eqnarray}}
\newcommand{\eea}{\end{eqnarray}}
\newcommand{\bml}{\begin{mathletters} \baselineskip 10pt}
\newcommand{\eml}{\baselineskip 12pt \end{mathletters}}

\def\lambdabar{\protect\@lambdabar}
\def\@lambdabar{%
\relax
\bgroup
\def\@tempa{\hbox{\raise.73\ht0
\hbox to0pt{\kern.2\wd0\vrule width.7\wd0
height.1pt depth.1pt\hss}\box0}}%
\mathchoice{\setbox0\hbox{$\displaystyle\lambda$}\@tempa}%
{\setbox0\hbox{$\textstyle\lambda$}\@tempa}%
{\setbox0\hbox{$\scriptstyle\lambda$}\@tempa}%
{\setbox0\hbox{$\scriptscriptstyle\lambda$}\@tempa}%
\egroup
}

\newcommand{\DL}{\delta \mathscr{L}}
\newcommand{\SCS}{\mathscr{S}}
\newcommand{\SCP}{\mathscr{P}}

\newcommand{\sfrac}[2]{{\textstyle \frac{#1}{#2}}}

\newcommand{\vcb}[1]{\mbox{\bf #1}}

\newcommand{\vc}[1]{\mbox{\boldmath$#1$}}

\begin{document}


\title{On the Observation of Vacuum Birefringence}

\author{Thomas Heinzl}
\email{theinzl@plymouth.ac.uk}

\affiliation{School of Mathematics and Statistics, University of
Plymouth\\
Drake Circus, Plymouth PL4 8AA, UK}



\author{Ben Liesfeld}

\author{Kay-Uwe Amthor}

\author{Heinrich Schwoerer}

\author{Roland Sauerbrey}
\email{sauerbrey@ioq.uni-jena.de}

\affiliation{Institut f{\"u}r Optik und Quantenelektronik,
Friedrich-Schiller-Universit{\"a}t Jena\\
Max-Wien-Platz 1, 07743 Jena, Germany}


\author{Andreas Wipf}%
\email{wipf@tpi.uni-jena.de}

\affiliation{Theoretisch-Physikalisches Institut,
Friedrich-Schiller-Universit\"at Jena\\
Max-Wien-Platz 1, 07743 Jena, Germany}


\date{\today}

\begin{abstract}
We suggest an experiment to observe vacuum birefringence induced
by intense laser fields. A high-intensity laser pulse is focused
to ultra-relativistic intensity and polarizes the vacuum which
then acts like a birefringent medium. The latter is probed by a
linearly polarized x-ray pulse. We calculate the resulting
ellipticity signal within strong-field QED assuming Gaussian
beams. The laser technology required for detecting the signal will
be available within the next three years.
\end{abstract}

\maketitle


\noindent
The interactions of light and matter are described by
quantum electrodynamics (QED), at present the best-established
theory in physics. The QED Lagrangian couples photons to charged
Dirac particles in a gauge invariant way. At photon energies small
compared to the electron mass, $\omega \ll m_e$, electrons (and
positrons) will generically not be produced as real particles.
Nevertheless, as already stated by Heisenberg and Euler, ``...even
in situations where the [photon] energy is not sufficient for
matter production, its virtual possibility will result in a
`polarization of the vacuum' and hence in an alteration of
Maxwell's equations'' \cite{heisenberg:1936}. These authors were
the first to explicitly derive the nonlinear terms induced by QED
for small photon energies but arbitrary intensities (see also
\cite{weisskopf:1936}).

The most spectacular process resulting from these modifications
presumably is pair production in a constant electric field. This
is an \textit{absorptive} process as photons disappear by
disintegration into matter pairs. It can occur for field strengths
larger than the critical one given by \cite{sauter:1931,schwinger:1951}
\be
  E_c \equiv \frac{m_e^2}{e} \simeq 1.3 \times 10^{18} \,
  \mathrm{V/m} \; .
\ee
In this electric field an electron gains an energy $m_e$ upon
travelling a distance equal to its Compton wavelength,
$\lambdabar_e = 1/m_e$. The associated intensity is $I_c = E_c^2
\simeq 4.4 \times 10^{29}$ W/cm$^2$ such that both field strength
and intensity are way out of experimental reach for the time being
--  unless one can utilize huge relativistic gamma factors
produced by large scale particle accelerators
\cite{bula:1996,burke:1997}.

Alternatively, there are also \textit{dispersive} effects that may
be considered. These include many of the phenomena studied in
nonlinear optics as well as ``birefringence of the vacuum'' first
addressed by Klein and Nigam \cite{klein:1964} in 1964, soon
followed by more systematic studies
\cite{BB:1967,bialynicka-birula:1970,brezin:1970b}. In essence,
the polarized QED vacuum acts like a birefringent medium (e.g.~a
calcite crystal) with two indices of refraction depending on the
polarization of the incoming light.  In a static magnetic field of
$5\, \mathrm{T}$ a light polarization rotation has recently been
observed \cite{zavattini:2005}. The measured signal differs from
the QED expectations and may be caused by a new coupling of
photons to an hitherto unobserved pseudoscalar.

Detection of the tiny dispersive effects is an enormous challenge.
In this paper we point out that several orders of magnitude in
field strength may be gained by employing high-power lasers.
Distorting the vacuum with lasers has been suggested long ago
\cite{brezin:1970b} but was not considered experimentally for lack
of sufficient laser power. However, recent progress in both laser
technology and x-ray detection has lead to novel experimental
capabilities. It is therefore due time to specifically address the
feasibility of a strong-field laser experiment to measure vacuum
birefringence. In the light of the results \cite{zavattini:2005}
such experiments are also necessary in order to test whether
strong electromagnetic fields provide windows into new physics.

We intend to utilize the high-repetition rate petawatt class laser
system POLARIS which is currently under construction at the Jena
high-intensity laser facility and which will be fully operational
in 2007 \cite{hein:2004}. POLARIS consists of a diode-pumped laser
system based on chirped pulse amplification (CPA) which will be
operating at $\Lambda = 1032\,\mathrm{nm}$ ($\Omega = 1.2$ eV)
with a repetition rate of $0.1\,\mathrm{Hz}$. A pulse duration of
about $140\,\mathrm{fs}$ and a pulse energy of $150\,\mathrm{J}$
in principle allows to generate intensities in the focal region of
$I=10^{22} \, \mathrm{W/cm^{2}}$. This corresponds to a
substantial electric field $E \simeq 2 \times 10^{14} \,
\mathrm{V/m}$, still about four orders of magnitude below $E_c$.

\begin{figure}
\begin{center}

\includegraphics[scale=0.33]{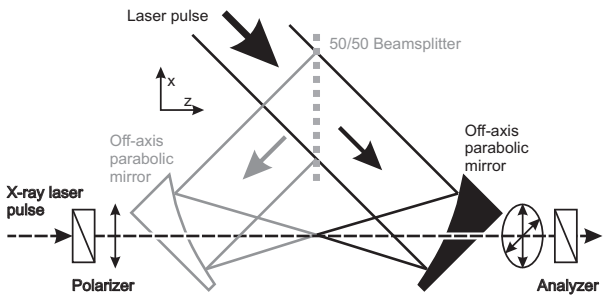}

\end{center}

\caption{\label{FIG:exp_setup}Proposed experimental setup for the
demonstration of vacuum birefringence: A high-intensity laser
pulse is focused by an $F/2.5$ off-axis parabolic mirror. A hole
is drilled into the parabolic mirror in alignment with the
$z$-axis (axes as indicated) in such a way that an x-ray pulse can
propagate along the $z$-axis through the focal region of the
high-intensity laser pulse. Using a polarizer-analyzer pair the
ellipticity of the x-ray pulse may be detected. Shown in grey:
Extension of the setup for the generation of counter propagating
laser pulses and a high-intensity standing wave which may be used
for pair creation.}
\end{figure}

The proposed experimental setup is shown in
Fig.~\ref{FIG:exp_setup}. A high-intensity laser pulse is focused
by an off-axis parabolic mirror.  A linearly polarized
laser-generated ultra-short x-ray pulse is aligned collinearly
with the focused optical laser pulse. After passing through the
focus the laser induced vacuum birefringence will lead to a small
ellipticity of the x-ray pulse which will be detected by a high
contrast x-ray polarimeter \cite{hart:1991}. The whole setup is
located in an ultra-high vacuum chamber and is entirely computer
controlled.

Shown in grey in Fig.~\ref{FIG:exp_setup} is an extension of the
setup which enables us to accurately overlap two counter
propagating high-intensity laser pulses. Accurate control over
spatial and temporal overlap was convincingly demonstrated
carrying out an autocorrelation of the laser pulses \emph{at full
intensity} \cite{liesfeld:2005} and generating Thomson
backscattered x-rays from laser-accelerated electrons
\cite{schwoerer:2005}. This counter propagating scheme, a
table-top ``photon collider'', may also be employed for pair
creation from the vacuum. For the x-ray probe pulse we have chosen
an x-ray source of photon energy $\omega \simeq 1$ keV, since the
birefringence signal is proportional to $\omega^2$ (see below) .
Our long-term plans are to replace the present source by an x-ray
free electron laser (XFEL) or by a laser-based Thomson
backscattering source \cite{schwoerer:2005} both of which deliver
ultrashort and highly polarized x-rays.

Refraction is a dispersive process based on modified propagation
properties of the probe photons travelling through a region where
a (strong) background field is present. The resulting corrections
to pure Maxwell theory to leading order in the probe field $a_\mu$
may be expressed in terms of an effective action
\cite{brezin:1970b}
\be \label{DELTA_S}
  \delta S \equiv \sfrac{1}{2} \int d^4 x \, d^4 y \, a_\mu (x)
  \Pi^{\mu \nu} (x, y; A) a_\nu(y) \; ,
\ee
where $A_\mu$ denotes the background field and $\Pi^{\mu\nu}$ the
polarization tensor. To lowest order in a loop (or $\hbar$)
expansion the former is given by the Feynman diagram
\be \label{POLFEYN}
  \includegraphics[scale=0.9]{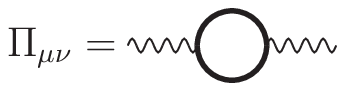}
\ee
with the heavy lines denoting the dressed propagator depending on
the background field $A$,
\be
  \includegraphics[scale=0.85]{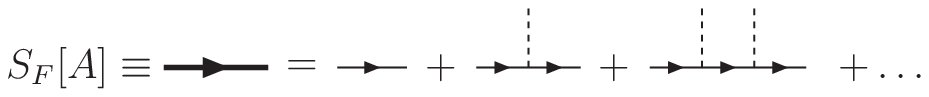}
\ee
Hence, $S_F[A]$ is an infinite series of diagrams where the $n$th
term corresponds to the absorption and/or emission of $n-1$
background photons (represented by the dashed external lines) by
the ``bare'' electron.

The dressed propagator is known exactly only for a few special
background configurations (see \cite{dittrich:2000} for an
overview). Typically, one obtains rather unwieldy integral
representations which have to be analyzed numerically. In our
case, however, we can exploit the fact that we are working in the
regime of both low energy and small intensities leading to
\textit{two} small parameters \cite{affleck:1987}, namely
\bea
  \nu^2 &\equiv& \omega^2 / m_e^2 \simeq 4 \times 10^{-6} \;,
  \label{XI} \\
  \epsilon^2 &\equiv& E^2/E_c^2 = I/I_c \simeq 2 \times 10^{-8} \;.
  \label{EPSILON} \eea
\textit{Low intensity}, $\epsilon^2 \ll 1$, means that we can work
to lowest nontrivial order in the external field i.e.\
$O(\epsilon^2)$. In terms of Feynman diagrams (\ref{POLFEYN}) then
reduces to
\be \label{POLEXPAND}
  \includegraphics[scale=0.85]{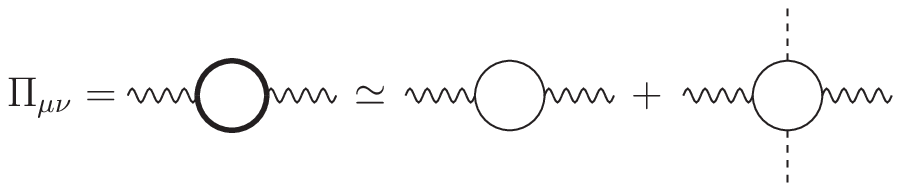}
\ee
%

\textit{Low energy}, $\nu \ll 1$, implies that we may safely
expand $\Pi^{\mu\nu}$ in derivatives or, after Fourier
transformation, in powers of the probe 4-momentum $k = \omega (1,
n \vcb{k})$ where $\vcb{k}^2 = 1$ and $n \ge 1$ is the index of
refraction. Thus, the derivative expansion is in powers of
$\omega^2$ or, equivalently, of $\nu^2$. Again we restrict our
analysis to leading order which turns out to be $\nu^2$. The first
vacuum polarization diagram in (\ref{POLEXPAND}) is $O(\nu^4)$
while the second is $O(\epsilon^2 \nu^2)$ so we may safely neglect
the former. The low-energy limit of the remaining diagram is
obtained from the celebrated Heisenberg-Euler Lagrangian
\cite{heisenberg:1936} which to leading order in $\epsilon^2$ is
given by
\be \label{LHE}
  \DL (\SCS, \SCP) = \sfrac{1}{2} \gamma_- \SCS^2 +
  \sfrac{1}{2} \gamma_+ \SCP^2  \; .
\ee
The basic building blocks in (\ref{LHE}) are the scalar and
pseudoscalar invariants
\cite{schwinger:1951,bialynicka-birula:1970}
\bea
  \SCS &\equiv& - \sfrac{1}{4} F_{\mu\nu} F^{\mu\nu} = \sfrac{1}{2}
  ( \vc{E}^2 - \vc{B}^2) \; , \label{S}
  \\
  \SCP &\equiv& - \sfrac{1}{4} F_{\mu\nu} \tilde{F}^{\mu\nu} = \vc{E} \cdot
  \vc{B} \label{P} \; ,
\eea
where $F_{\mu\nu}$ denotes the electromagnetic field-strength
tensor (comprising both background and probe photon field) and
$\tilde{F}_{\mu\nu}$ its dual. The nonlinear couplings in
(\ref{LHE}) are given by
\be \label{GAMMA_PM}
  \gamma_+ \equiv 7 \rho \; , \quad \gamma_- \equiv 4 \rho \; ,
  \quad \rho \equiv \frac{\alpha}{45 \pi} \frac{1}{E_c^2} \; ,
\ee
with $\alpha = 1/137$ being the fine-structure constant.

To proceed we split the fields into an intense (laser) background
and a weak probe field according to the replacement $F_{\mu\nu}
\to F_{\mu\nu} + f_{\mu\nu}$ with upper (lower) case letters for
electromagnetic quantities henceforth referring to the background
(probe). In the following we regard the plane wave probe field
$f_{\mu\nu}$ as a weak disturbance on top of the strong background
field $F_{\mu\nu}$ which we take as an electromagnetic wave of
frequency $\Omega$. It can be a plane or standing wave or more
realistic variants thereof like Gaussian beams (see discussion
below). In any case, for the actual experiment we will have the
hierarchy of frequencies $\Omega \ll \omega \ll m_e$ in agreement
with (\ref{XI}).

The leading-order contribution to the polarization tensor is found
by performing the split $F \to F + f$ in the Heisenberg-Euler
action, $\delta S = \int d^4 x \, \DL$, and writing it in the form
(\ref{DELTA_S}). This yields a polarization tensor
\be \label{PI}
  \Pi^{\mu\nu} = - \gamma_- \, k^2 \SCS \, \mathbb{P}^{\mu\nu} + \gamma_-
  \, b^\mu b^\nu + \gamma_+ \, \tilde{b}^\mu \tilde{b}^\nu \; ,
\ee
where $\mathbb{P}^{\mu\nu} = g^{\mu\nu} - k^\mu k^\nu / k^2$ is
the standard projection orthogonal to $k$ and $\SCS$ denotes the
\textit{background} invariant. In addition we have introduced the
new 4-vectors \cite{bialynicka-birula:1970},
\be \label{BBT}
  b^\mu \equiv F^{\mu\nu}k_\nu \; , \quad \tilde{b}^\mu \equiv
  \tilde{F}^{\mu\nu} k_\nu \; .
\ee
Note that we have $b \cdot k = 0 = \tilde{b} \cdot k$ and hence
$\Pi^{\mu\nu}k_\nu = 0$ as required by gauge invariance. It is
useful to diagonalize $\Pi^{\mu\nu}$ and rewrite it in terms of a
spectral decomposition. In full generality this is a bit awkward,
but for our purposes matters can be simplified. The eigenvalues of
$\Pi^{\mu\nu}$ in principle depend on the four invariants $k^2$,
$\SCS$, $\SCP$ and $b^2$. From (\ref{PI}) we note that there is no
$\SCP$ dependence and that only the combination $k^2 \SCS$
appears. Let us count powers of $\epsilon$ and $\nu$ to determine
the relative magnitudes of the invariants. If we write the index
of refraction as $n = 1 + \Delta$ we expect $\Delta =
O(\epsilon^2)$ the deviation of $n$ from unity being due to the
external fields. Hence $k^2$ is no longer zero but rather $k^2 =
O(\epsilon^2 \nu^2)$ implying $k^2 \SCS = O(\epsilon^4 \nu^2)$.
For generic geometrical settings (see below) the invariant $b^2 =
O(\epsilon^2 \nu^2)$. The upshot of this power counting exercise
is the important inequality
\be \label{INEQ}
  |k^2 \SCS| \ll |b^2| \; ,
\ee
by means of which we may neglect $k^2 \SCS$. This justifies the
statement in \cite{dittrich:2000} that to leading order in
$\epsilon$ and $\nu$ the eigenvalues of $\Pi^{\mu\nu}$ do not
depend on the invariants $\SCS$ and $\SCP$. Hence, under the
assertion (\ref{INEQ}) constant fields behave as \textit{crossed
fields} ($\vc{E}$ and $\vc{B}$ orthogonal and of the same
magnitude) for which strictly $\SCS = \SCP = 0$. In addition, one
has $b^2 = \tilde{b}^2$ and $b \cdot \tilde{b} = 0$ so that
(\ref{PI}) turns into the spectral representation
\be \label{PI_DIAG}
  \Pi^{\mu\nu} = \gamma_- \, b^\mu b^\nu + \gamma_+ \, \tilde{b}^\mu
  \tilde{b}^\nu \; .
\ee
We read off that the (nontrivial) eigenvectors are given by
(\ref{BBT}) corresponding to eigenvalues $\gamma_\pm b^2 (k)$.
Note that $b^2$ is the only nonvanishing invariant which can be
built from crossed fields.

Adopting a plane wave ansatz for the probe field $a_\mu$ yields a
homogeneous wave equation which in momentum space becomes linear
algebraic. It has nontrivial solutions only if a secular equation
holds which determines the dispersion relations for $k^2$. With
the eigenvalues given above there are \textit{two} of them, $k^2 -
\gamma_\pm b^2 (k) = 0$. Inserting $k = \omega (1, n \vcb{k})$, we
finally obtain two solutions for the index of refraction,
\be \label{N1}
  n_\pm = 1 + \sfrac{1}{2} \gamma_\pm Q^2 \; .
\ee
The nonnegative quantity $Q^2$ is an energy density which in
3-vector notation becomes
\be
  Q^2 = \vc{E}^2 + \vc{B}^2 - 2 \vc{S} \cdot \vcb{k} - (\vc{E} \cdot
  \vcb{k})^2 - (\vc{B} \cdot \vcb{k})^2 \; ,
\ee
with $\vc{S} = \vc{E} \times \vc{B}$ being the Poynting vector. The
inequality (\ref{INEQ}) holds as long as $Q^2 \ne 0$. The indices of
refraction become maximal if probe and background are counter
propagating (`head-on collision'), $\vcb{k} = - \vc{S}/|\vc{S}|$,
whereupon
\be \label{QQ}
  Q^2 = \vc{E}^2 + \vc{B}^2 + 2 |\vc{S}| \equiv 4 I \; ,
\ee
with $I$ denoting the background intensity. Note that one gains a
factor of four as compared to a purely electric or purely magnetic
background. Plugging (\ref{QQ}) into (\ref{N1}) the indices of
refraction become $n_\pm = 1 + 2 \gamma_\pm I$ or, upon inserting
$\gamma_\pm$,
\be
  n_\pm = 1 + \left\{14 \atop 8 \right\} \rho I = 1 +
  \frac{\alpha}{45\pi} \left\{14 \atop 8 \right\}
  \frac{I}{I_c}  \; .
\ee
To the best of our knowledge, these values have first been
obtained in \cite{BB:1967}. They imply birefringence with a
relative phase shift between the two rays proportional to
$\triangle n \equiv n_+ - n_-$,
\be \label{DPHI}
  \triangle \phi = 2\pi \frac{d}{\lambda} \triangle n = \frac{4
  \alpha}{15} \frac{d}{\lambda} \frac{I}{I_c} = \frac{4
  \alpha}{15} \frac{d}{\lambda} \epsilon^2 \; .
\ee

\begin{table}
\caption{\label{TABLE:1} Numerical values for the phase shift
(\protect\ref{DPHI_KAPPA}) and ellipticity signal $\delta^2$.
First line: present specifications of the Jena laser facility.
Second line: optimal scenario with XFEL probe and large Rayleigh
length. The peak intensity is taken to be $I_0 = 10^{22}$
W/cm$^2$.}
\begin{ruledtabular}
\begin{tabular}{rrccc}
$\omega$ / keV & $\lambda$ / nm & $z_0$ / $\mu$m &$\triangle \phi$ / rad & $\delta^2$\\
\hline\\[-5pt]
1.0  & 1.2   & 10 & $1.2 \times 10^{-6}$ & $3.4 \times 10^{-13}$ \\
15   & 0.08  & 25 & $4.4 \times 10^{-5}$ & $4.8 \times 10^{-10}$
\end{tabular}
\end{ruledtabular}
\end{table}
%

A realistic laser field will lead to an intensity distribution
along the $z$-axis (choosing $\vcb{k} = \vc{e}_z$). If $z_0$
measures the typical extension of the distribution we may set $s
\equiv z/z_0$ and write the intensity as $I(s) = I_0 \, g(s)$ with
peak intensity $I_0$ and a dimensionless distribution function
$g(s)$. The phase shift (\ref{DPHI}) is then replaced by the
expression \cite{koch:2004}
\be \label{DPHI_KAPPA}
  \triangle \phi = \frac{4
  \alpha}{15} \frac{z_0}{\lambda} \frac{I_0}{I_c} \kappa \; ,
\ee
where the correction factor $\kappa$ is the integral
\be
  \kappa = \kappa(s_0) \equiv \int_{-s_0}^{s_0} ds \, g(s) = O(1)
  \;  .
\ee
Here, $s_0$ denotes the half-width of the intensity distribution
in units of $z_0$. In general it is a reasonable approximation to
let $s_0 \to \infty$. For a \textit{single} Gaussian beam, $z_0$
is the Rayleigh length and the intensity $I_1$ follows a Lorentz
curve, hence $g_1(s) = 1/(1+s^2)$ implying $\kappa_1 (\infty) =
\pi$. Identifying $d=2z_0$ this differs from (\ref{DPHI}) by a
factor of $\pi/2 = O(1)$. For \textit{two} counter propagating
Gaussian beams (`standing wave') obtained from splitting a beam of
intensity $I_1$ one gains a factor of two in peak intensity but
the distribution gets thinned out due to the usual $\cos^2$
modulation, which cancels the gain in intensity leading to the
\textit{same} correction factor $\kappa_2 = \pi = \kappa_1$.

A linearly polarized electromagnetic wave undergoing vacuum
birefringence with a polarization vector oriented under an angle
of $45^\circ$ with respect to both background fields $\vc{E}$ and
$\vc{B}$ will be rendered elliptically polarized with ellipticity
$\delta$ (ratio of the field vectors). In the experiment,
intensities will be measured and the experimental quantity to be
determined is $\delta^2 \simeq (\frac{1}{2} \Delta \phi)^2 $. In
Table~\ref{TABLE:1} expected ellipticity values for given
experimental parameters are listed.

These results clearly show the challenging nature but also the
feasibility of the proposed experiment. Presently, a petawatt
class laser facility such as POLARIS is expected to reach about
$10^{22}\,\mathrm{W/cm^2}$ at unprecedented repetition rates of
$\sim 0.1\,\mathrm{Hz}$ \cite{hein:2004}. The values of $\delta^2$
obtained for such lasers (Tab.~\ref{TABLE:1}) are at the limit of
the accuracy that can now be obtained with high-contrast x-ray
polarimeters using multiple Bragg reflections from channel-cut
perfect crystals \cite{hart:1991,hasegawa:1999,alp:2000}. These
instruments are in principle capable of a sensitivity of $\delta^2
\simeq 10^{-11}$ \cite{alp:2000}. Since the expected signal is
proportional to both $I^2$ and $\lambda^{-2}$ it may be greatly
enhanced by increasing the laser intensity or choosing a smaller
probe pulse wave length. For example, with the proposed ELI laser
facility reaching $10^{25}\,\mathrm{W/cm^2}$ \cite{mourou:2005} a
sensitivity of the polarimeter of only $10^{-7} \dots 10^{-4}$ is
required which is within presently demonstrated values of
sensitivity \cite{hart:1991}. The required x-ray probe pulse may
be generated either with an XFEL synchronized to a petawatt laser
or by the use of Thomson scattered laser photons from
monochromatic laser accelerated electron beams
\cite{schwoerer:2005,faure:2004}.

It seems worthwhile to point out that although a standing wave for
the background (which may be created in the ``photon collider''
setup as shown in Fig.~\ref{FIG:exp_setup}) does not lead to an
increase in integrated intensity and hence of the birefringence
signal, it \textit{does} yield double peak intensity. This is
important for the observation of effects sensitive to localized
intensity like Cherenkov radiation and pair production.

\noindent This work was supported by the DFG Project
TR~18. The authors gratefully acknowledge stimulating discussions
with H.~Gies, A.~Khvedelidze, M.~Lavelle, V.~Malka, D.~McMullan,
G.~Mourou, A.~Nazarkin, A.~Ringwald and O.~Schr{\"oder}.


\bibliographystyle{../../bibfiles/h-physrev}
\bibliography{../../bibfiles/laser}

\begin{thebibliography}{20}
\bibitem{heisenberg:1936}
W.~Heisenberg and H.~Euler, Z.~Phys.\ \textbf{98}, 714 (1936).
\bibitem{weisskopf:1936}
V.~Weisskopf, K.~Dan.~Vidensk.~Selsk.~Mat.~Fys.~Medd.,
\textbf{14}, 6 (1936), reprinted in \textit{Quantum
Electrodynamics}, J.~Schwinger, ed., Dover, New York 1958.
\bibitem{sauter:1931}
F.~Sauter, Z.~Phys.\ \textbf{69}, 742 (1931).
\bibitem{schwinger:1951}
J.~Schwinger, Phys.\ Rev.\ \textbf{82}, 664 (1951).
\bibitem{bula:1996}
C.~Bula et~al., Phys.\ Rev.\ Lett.\ \textbf{76}, 3116 (1996).
\bibitem{burke:1997}
D.~Burke et~al., Phys.\ Rev.\ Lett.\ \textbf{79}, 1629 (1997).
\bibitem{klein:1964}
J.~Klein and B.~Nigam, Phys.\ Rev.\ \textbf{135}, B1279 (1964).
\bibitem{BB:1967}
R.~Baier and P.~Breitenlohner, Acta Phys.~Austriaca \textbf{25}, 212
(1967);  Nuovo Cim.~B \textbf{47}, 117 (1967).
\bibitem{bialynicka-birula:1970}
Z.~Bia{\l}ynicka-Birula and I.~Bia{\l}ynicki-Birula, Phys.\ Rev.\
D\textbf{2}, 2341 (1970).
\bibitem{brezin:1970b}
E.~Brezin and C.~Itzykson, Phys.\ Rev.\ D\textbf{3}, 618 (1970).
\bibitem{zavattini:2005}
E.~Zavattini et.~al., PVLAS collaboration (2005), hep-ex/0507107.
\bibitem{hein:2004}
J.~Hein et al., Appl.\ Phys.\ B \textbf{79}, 419 (2004).
\bibitem{hart:1991}
M.~Hart et al., Rev.~Sci.~Instrum.\ \textbf{62}, 2540 (1991).
\bibitem{liesfeld:2005}
B.~Liesfeld et al., Appl.~Phys.~Lett., \textbf{86}, 161107 (2005).
\bibitem{schwoerer:2005}
H.~Schwoerer et al., Phys.~Rev.~Lett.\ (2005), accepted for
publication.
\bibitem{dittrich:2000}
W.~Dittrich and H.~Gies, \textit{Probing the quantum vacuum},
vol.\ 166 of  \textit{Springer Tracts Mod. Phys.} (Springer,
Berlin, 2000).
\bibitem{affleck:1987}
I.~Affleck and L.~Kruglyak, Phys.\ Rev.\ Lett.\ \textbf{59}, 1065
(1987).
\bibitem{koch:2004}
K.~Koch, diploma thesis, Jena (2004), in German.
\bibitem{hasegawa:1999}
Y.~Hasegawa et al., Acta Cryst.~A \textbf{55}, 955 (1999).
\bibitem{alp:2000}
E.~Alp, W.~Sturhahn, and T.~Toellner, Hyperfine Interactions
\textbf{125}, 45 (2000).
\bibitem{mourou:2005}
G.~Mourou and V.~Malka, private communication (2005).
\bibitem{faure:2004}
J.~Faure et al., Nature \textbf{431} (7008), 541 (2004).
\end{thebibliography}

\end{document}